Version 1.1

# Long-Term Trends in Personal Given Name Frequencies in England and Wales[1]

Douglas A. Galbi
Senior Economist
Federal Communications Commission[2]

July 20, 2002

*Abstract*

The frequency distribution of personal given names offers important evidence about the information economy.  This paper presents data on the popularity of the most frequent personal given names (first names) in England and Wales over the past millennium.  The popularity of a name is its frequency relative to the total name instances sampled.  The data show that the popularity distribution of names, like the popularity of other symbols and artifacts associated with the information economy, can be helpfully viewed as a power law.  Moreover, the data on name popularity suggest that historically distinctive changes in the information economy occurred in conjunction with the Industrial Revolution.

The symbolic choices of ordinary persons are becoming more extensive and more significant. New information and communications technologies provide new capabilities for manipulating bits of information. These technologies give ordinary persons in their daily lives more extensive opportunities to choose among symbols, to re-arrange symbols, and to share symbols. Human work is becoming less like farming and more like singing. The information economy – the way persons make symbolic choices and the aggregate effects of such choices – increasingly matters for economic growth, industrial organization, and social life.

The frequency of personal given names provides important historical evidence about the information economy. Persons who have the same given name literally share the experience of being called by that name. The frequency distribution of names thus indicates an aspect of shared symbolic experience. Choosing a good name involves assessing the social valuation of a name. The frequency distribution of names thus provides evidence about symbolic valuation. Naming is often governed by norms, such as naming after parents, godparents, biblical figures, or deceased siblings. These norms – common laws in the economy of names – evolve through awareness of patterns of cases and possibilities for differences and exceptions. They too are aspects of the information economy. More abstractly, the frequency distribution of personal names, graphed as the logarithm of name popularity against the logarithm of name popularity rank, looks similar to other frequency or popularity distributions where persons and organizations are free to create and choose among many collections of symbols instantiated and used in a similar way. Thus naming appears to be representative of patterns of behavior in the information economy.

This paper shows that, since early in the nineteenth century, the frequency distribution of personal given names in England has evolved differently than it did over the previous eight centuries. Simple indicators of this change are the trend in the popularity (frequency relative to the total number of names in the sample) of the most frequent names. The popularity of the most frequent name, the three most frequent names, and the ten most frequent names show no trend from circa 1300 to 1800. Since then all these measures have dropped dramatically. This latter development reflects a "flattening" in the name frequency distribution, viewed as a graph of the logarithm of name popularity against the logarithm of name popularity rank. This change in the evolution of the name frequency distribution early in the nineteenth century may indicate a more general change in the information economy about that time.

## I. Popularity of the Most Frequent Personal Given Names

This section provides data on the popularity of the most frequency personal given names in England over the past millennium. Measuring name frequencies in actual samples requires attention to name definition and standardization. Given names can include multiple names and name variants, as well as abbreviations, non-standard spellings, and mistakes in recording. Unlike sampling variability, coding variability does not fall with sample size. Throughout the analysis in this paper, names have been truncated to the shorter of either the first eight letters of the given name or the letters preceding the first period, space, hyphen, or other non-alphabetic character. These shortened names have then been standardized through a name coding available on the Internet for public inspection, use, and improvement on an open source basis.[3] This procedure attempts to identify feasibly and consistently names with common communicative properties.[4] Experience with different name samples suggests that this procedure can reduce coding variability to less that half a percentage point for the popularity of a single name and less than three percentage points for total popularity of the top ten names (Galbi 2001, Sec. I.B. and Appendix B).

Over the past two hundred years, the popularity of the most frequent personal given names in England and Wales has steadily declined. Table 1 shows popularity statistics for the most frequent names. In England and Wales from 1800 to 1994, the popularity of the most frequent female and male names fell from 23.9% and 21.5% to 3.4% and 4.2%, respectively. The popularity of the ten most frequent names for females and males fell from 82.0% and 84.7% to 23.8% and 28.4%, respectively.

| | Table 1 Popularity of Personal Given Names: England and Wales 1800 to 1994 | | | | | | |
|---|---|---|---|---|---|---|---|
| | Females | | | | Males | | |
| Birth Year | Top Name Name | Top Name Pop. | Top 3 Pop. | Top 10 Pop. | Top Name Name | Top 3 Pop. | Top 10 Pop. |
| 1800 | Mary | 23.9% | 53.2% | 82.0% | John | 21.5% | 51.5% | 84.7% |
| 1810 | Mary | 22.2% | 50.7% | 79.4% | John | 19.0% | 47.0% | 81.4% |
| 1820 | Mary | 20.4% | 47.7% | 76.5% | John | 17.8% | 44.9% | 80.4% |
| 1830 | Mary | 19.6% | 45.4% | 75.8% | John | 16.4% | 42.3% | 78.2% |
| 1840 | Mary | 18.7% | 43.2% | 75.0% | William | 15.4% | 40.3% | 76.0% |
| 1850 | Mary | 18.0% | 41.0% | 72.1% | William | 15.2% | 38.7% | 73.8% |
| 1860 | Mary | 16.3% | 37.0% | 68.3% | William | 14.5% | 36.2% | 69.8% |
| 1870 | Mary | 13.3% | 31.5% | 61.1% | William | 13.1% | 31.7% | 63.5% |
| 1880 | Mary | 10.6% | 25.4% | 53.8% | William | 11.7% | 28.5% | 58.9% |
| 1900 | Elizabet | 7.2% | 16.2% | 38.5% | William | 9.0% | 22.9% | 50.9% |
| 1925 | Mary | 6.7% | 16.8% | 38.7% | John | 7.3% | 17.6% | 38.0% |
| 1944 | Margaret | 4.5% | 12.6% | 31.7% | John | 8.3% | 20.7% | 39.9% |
| 1954 | Susan | 6.1% | 13.2% | 32.5% | David | 6.3% | 17.4% | 37.8% |
| 1964 | Susan | 3.6% | 10.3% | 28.6% | Paul | 5.6% | 15.9% | 39.4% |
| 1974 | Sarah | 4.9% | 12.3% | 28.0% | Mark | 4.6% | 12.5% | 33.1% |
| 1984 | Sarah | 4.1% | 11.0% | 27.3% | James | 4.3% | 11.8% | 32.3% |
| 1994 | Emily | 3.4% | 8.6% | 23.8% | James | 4.2% | 11.0% | 28.4% |

Note: Based on Galbi (2001), Table 3, p 15 and underlying data. See Appendix D of that paper for sources and details of analysis..

Prior to the beginning of the nineteenth century, the popularity of the most frequent personal given names in England was higher and more stable. Tables 2 and 3 provide evidence on name popularity from late in the eleventh century through early in the nineteenth century. From 1300 to 1800, popularities of 20%, 50%, and 80% seem to be roughly representative figures for the top name, top three names, and top ten names for both females and males. As Table 1 shows, the corresponding figures for the late twentieth century are much lower – about 4%, 10%, and 25%.

Table 2
Popularity of Personal Given Names in England before 1825

| Year, Location | Females | | | | | Males | | | | |
|---|---|---|---|---|---|---|---|---|---|---|
| | Top Name | Pop. | Top 3 Pop. | Top 10 Pop. | Sample Size | Top Name | Pop. | Top 3 Pop. | Top 10 Pop. | Sample Size |
| 1080, Winchester 1) | | | | | | Robert | 6.6% | 18.0% | 35% | 228 |
| 1120, Winchester 1) | | | | | | William | 6.6% | 15.8% | 30% | 912 |
| 1180, Winchester 1) | | | | | | William | 10.2% | 29.2% | 57% | 383 |
| 1200, Essex 2) | Alice | 11.3% | 27.4% | 56% | c. 1400 | William | 12.4% | 30.5% | 61% | c. 4000 |
| 1210, South 3) | Matilda | 16.2% | 39.9% | 70% | 173 | William | 14.4% | 32.7% | 65% | 877 |
| 1270, Rutland 4) | Alice | 19.4% | 51.0% | 84% | 206 | William | 15.2% | 35.6% | 76% | 1627 |
| 1300, Lincoln 5) | Alice | 17.1% | 42.4% | 75% | 1213 | John | 22.7% | 52.2% | 79% | 9390 |
| 1260, London 6) | | | | | | John | 17.6% | 39.7% | 69% | 814 |
| 1290, London 6) | | | | | | John | 23.3% | 44.8% | 73% | 1852 |
| 1510, London 7) | | | | | | John | 24.4% | 49.4% | 74% | 427 |
| 1610, London 7) | | | | | | John | 21.0% | 43.8% | 72% | 463 |
| 1825, London 8) | Mary | 19.2% | 43.9% | 73% | 63809 | William | 16.3% | 39.2% | 80% | 48275 |
| 1350, Manchester 9c) | | | | | | John | 22.7% | 47.4% | 92% | 717 |
| 1610, Manchester 10) | | | | | | John | 18.6% | 37.6% | 77% | 1298 |
| 1805, Manchester 11) | Mary | 25.8% | 48.2% | 84% | 1866 | John | 21.7% | 48.7% | 81% | 1935 |
| 1350, Yorkshire 9d) | Alice | 22.4% | 50.4% | 86% | 1794 | John | 33.5% | 66.8% | 93% | 1665 |
| 1620, Yorkshire 12) | Ann | 24.0% | 54.7% | 88% | 342 | John | 16.2% | 47.1% | 86% | 427 |
| 1670, Yorkshire 12) | Ann | 21.5% | 59.2% | 79% | 228 | William | 18.7% | 47.4% | 78% | 283 |
| 1720, Yorkshire 12) | Mary | 25.7% | 57.4% | 87% | 413 | John | 25.5% | 57.8% | 86% | 377 |
| 1770, Yorkshire 12) | Mary | 22.8% | 45.9% | 84% | 381 | John | 25.6% | 55.7% | 86% | 433 |
| 1825, Yorkshire 8) | Mary | 20.1% | 45.8% | 81% | 99299 | John | 18.8% | 44.2% | 79% | 91111 |

**Table 2 (cont'd)**

| | Females | | | | | Males | | | |
|---|---|---|---|---|---|---|---|---|---|
| Year, Location | Top Name | Pop. | Top 3 Pop. | Top 10 Pop. | Sample Size | Top Name | Pop. | Top 3 Pop. | Top 10 Pop. | Sample Size |
| 1350, North/Cumbria 9a) | | | | | | John | 34.5% | 64.6% | 89% | 328 |
| 1530, North/Cumbria 13) | Jane | 16.0% | 44.8% | 84% | 852 | John | 23.1% | 46.1% | 74% | 870 |
| 1550, North/Cumbria 13) | Margaret | 15.6% | 45.1% | 86% | 1491 | John | 21.7% | 44.4% | 75% | 1515 |
| 1580, North/Cumbria 13) | Margaret | 16.8% | 44.9% | 84% | 3750 | John | 18.0% | 39.4% | 71% | 3765 |
| 1610, North/Cumbria 13) | Elizabet | 15.8% | 43.8% | 84% | 4000 | John | 18.2% | 42.4% | 74% | 4044 |
| 1640, North/Cumbria 13) | Elizabet | 16.6% | 46.0% | 87% | 2888 | John | 19.7% | 46.7% | 75% | 2914 |
| 1670, North/Cumbria 13) | Elizabet | 16.5% | 45.1% | 86% | 3813 | John | 19.6% | 46.7% | 75% | 3834 |
| 1700, North/Cumbria 13) | Ann | 16.4% | 47.1% | 86% | 3064 | John | 21.1% | 49.5% | 77% | 3070 |
| 1730, North/Cumbria 13) | Ann | 18.1% | 50.0% | 87% | 2038 | John | 21.6% | 49.9% | 80% | 2038 |
| 1760, North/Cumbria 13) | Ann | 18.8% | 52.1% | 89% | 2830 | John | 23.2% | 51.4% | 81% | 2830 |
| 1790, North/Cumbria 13) | Mary | 19.4% | 50.8% | 89% | 2139 | John | 23.4% | 52.9% | 83% | 2141 |
| 1825, North/Cumbria 8) | Mary | 20.3% | 46.7% | 88% | 24857 | John | 21.8% | 49.9% | 85% | 21966 |
| 1350, Hereford 9b) | Alice | 21.9% | 47.2% | 84% | 576 | John | 34.8% | 58.9% | 89% | 2066 |
| 1700, Hereford 14) | | | | | | John | 20.3% | 49.9% | 78% | 931 |
| 1825, Hereford 8) | Mary | 21.7% | 56.1% | 85% | 6832 | John | 18.9% | 51.1% | 90% | 6350 |
| 1280, East Anglia 15) | | | | | | John | 22.3% | 47.1% | 74% | 391 |
| 1400, East Anglia 15) | | | | | | John | 36.1% | 63.3% | 90% | 590 |
| 1385, soldiers 19) | | | | | | John | 28.1% | 58.3% | 84% | 829 |
| 1550, sailors 16) | | | | | | John | 21.4% | 40.8% | 70% | 583 |
| 1560, Canterbury 17) | Elizabet | 13.6% | 33.6% | 74% | 661 | John | 20.3% | 46.9% | 75% | 5986 |
| 1560, Gloucester 18) | Joan | 18.7% | 45.5% | 88% | c. 4000 | John | 21.4% | 52.5% | 80% | c. 4000 |

Note: Adapted from Galbi (2001) Table 5 p. 20. For details and sources, see ibid p. 41-43, and Appendix D.

| Table 3 Personal Given Names in England: 1570 to 1700 |||
| --- | --- | --- |
| | Popularity of Top 3 Names ||
| Birth Years | Females | Males |
| 1570-1579 | 41.0% | 48.5% |
| 1580-1589 | 36.2% | 47.3% |
| 1590-1599 | 41.1% | 50.6% |
| 1600-1609 | 38.2% | 48.8% |
| 1610-1619 | 38.8% | 49.9% |
| 1620-1629 | 41.3% | 49.3% |
| 1630-1639 | 45.1% | 48.5% |
| 1640-1649 | 46.7% | 49.3% |
| 1650-1659 | 50.1% | 49.0% |
| 1660-1669 | 47.5% | 48.0% |
| 1670-1679 | 50.3% | 50.3% |
| 1680-1689 | 51.7% | 49.2% |
| 1690-1700 | 52.1% | 51.2% |
| Source: Smith-Banister (1997) Table 7.8, p. 150. Figures for "mean" across English regions are given above. Sample sizes and weighting not reported. |||

Significant changes prior to 1800 seemed to have a relatively small effect on the pattern of name popularity. The Norman Conquest of England in 1066 brought about an almost complete change in given names. Within a few generations, most persons used given names brought by the invaders. By about 1250, pre-Conquest names had essentially died out.[5] The influx of new names and the shift to them must have decreased the popularity of the most popular names until the new naming practices were well established throughout society. Yet only about a hundred years after the Conquest the popularity of the most popular male names in Winchester had risen to a level closer to that in 1300 and 1800 than that in the late twentieth century. The information economy of the twelfth century supported an astonishing capacity for dissemination of information (new names) and creation of social information (the pattern of popular names). The relatively high name popularity prior to 1800 was not simply an artifact of inertia created by naming norms or underdevelopment of the information economy.

## II. A Better View of Changes over the Past Two Hundred Years

Changes in the popularity of the most frequent name, most frequent three names, and most frequent ten names are part of a larger order of change that can be easily recognized graphically. Scholars analyzing personal given names have recognized that graphs of name frequencies have a characteristic shape (Eschel, 2001; Tucker, 2001; Galbi, 2001). The graph of the logarithm of name popularity against the logarithm of name popularity rank is the same as a graph of name frequencies, except that the left axis is labeled is more easily understood units.[6] The graph typically is nearly a straight line. This type of empirical regularity is called a power law. It describes the relative frequency or popularity of names. Hence a power law describes a relationship between the popularity of the most frequent name, the three most frequent names, and the ten most frequent names

Over the past two hundred years, the change in the popularity of the most frequent names has been associated with a flattening of the power law that best describes the name popularity distribution. Charts 1 and 2 show these graphs for names of females and males born in England and Wales in 1819-30 and in 1994. In both cases the slope of a line approximating the graph has become less negative. This means that the relative name popularities have become more equal. This change can be interpreted as a reduction in the magnitude of information encoded in the name distribution and an increase in the extent of personalization in naming (Galbi 2001, Sec. II.B).

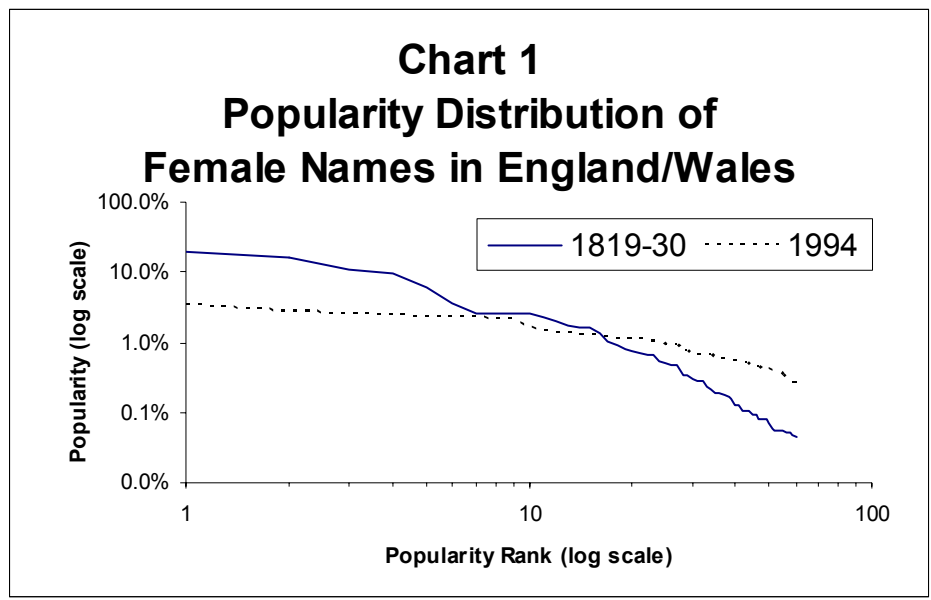

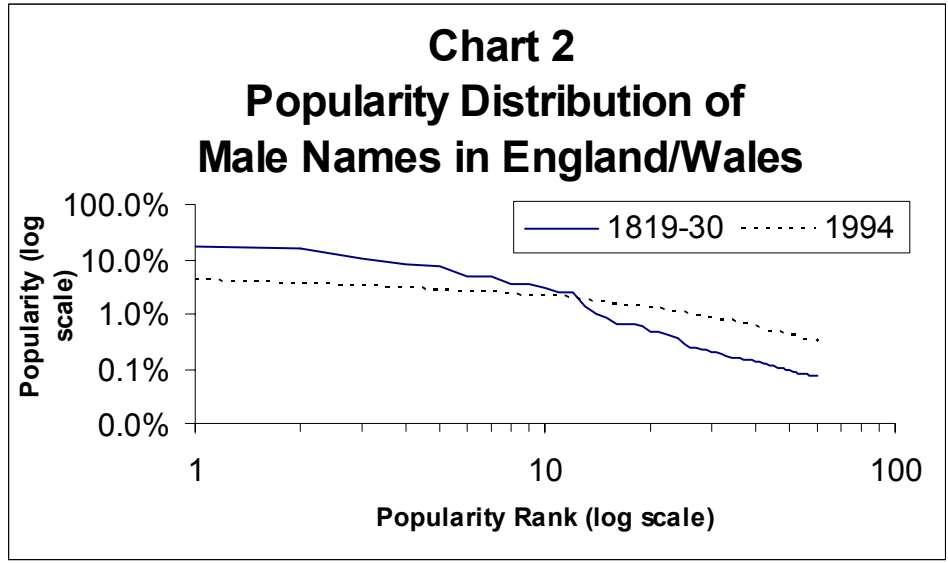

Empirical regularities such as those in Charts 1 and 2 are in fact prevalent in the information economy. Where persons and organizations are free to create and choose among many collections of symbols instantiated and used in a similar way, the relative popularity of the symbolic artifacts typically follows a power law. Thus the circulation of magazines of similar type has followed power laws throughout the twentieth century. The total box office receipts of movies follow a power law. The popularity of musical groups, as measured by "gold records," follows a power law (Chung and Cox, 1994). The popularity of Internet web sites, measured in users or page views, also follows a power law (Adamic and Huberman, 2000). Insights into the evolution of such power laws over time from study of names might contribute to more general insights into personal preferences, media diversity, information industry structure, and other aspects of the information economy.

**III. Understanding the Changes**

Although recent work on personal given names in England has emphasized name-sharing practices for understanding the frequency distribution of given names (Smith-Bannister, 1997), name-sharing practices have little direct relationship to the frequency distribution of names. Naming a significant share of children after parents, or after godparents, are equally consistent with a high or low popularity of the most frequent names. Similarly, having names freely chosen, i.e. chosen in absence of norms giving high value to the name of a person in a specific social position in relation to the person to be named, could produce high or low popularity of the most frequent names. The most that can be said for name-sharing is that a norm of naming after parents creates additional inertia in name popularity. Name popularity and its long-term evolution depend on factors other than name-sharing. The evolution of the name frequency distribution over time is a complicated dynamic system. Such systems can, in some circumstances, be highly sensitive to a particular factor, while in other circumstances, be totally unaffected by that factor. Moreover, boundary conditions, such as a small share of naming done in violation of prevailing norms, can determine the over-all state of the system.[7]

Analysis of long-term trends in personal given names in England and Wales suggests that significant changes in the information economy occurred in conjunction with the broad social and economic changes called the Industrial Revolution. The Industrial Revolution is associated with more rapid growth in population. The population of England in 1800 was about 50% greater than in 1300, while its population in 2000 was about six times greater than in 1800. The Industrial Revolution is also associated with much more rapid growth in income: real economic income per person probably increased by about a factor of four from 1300 to 1800, and by about a factor of 100 from 1800 to 2000.[8] But populations of much different sizes show similar naming patterns (Eshel (2001), Galbi (2001) Table 4), and it is not clear how the level of income itself would effect naming. The Industrial Revolution also produced major changes in social networks and the social context of personal activity. These changes probably drove changes in the information economy.

Whether information and communication technologies have created or will create a "new economy" is an important public policy issue. These technologies enable persons to interact in new ways that may be as significant as the changes associated with the Industrial Revolution. Consider, for example, the creation of knowledge about aggregate patterns of personal given names. Large compilations of name frequencies can be easily shared on the Internet. I have benefited from such sharing of information in writing this paper, and I have made much more extensive data on name frequencies than can be reported in this paper freely available on the Internet.[9] If other scholars use the Internet in similar ways, this sub-field of onomastics could develop much more rapidly than it has in the past. The same might be true of many other areas of activity. Analyzing the popularity distribution of personal given names thus offers a particularly rich means for understanding changes in the information economy.

# Appendix – Personal Given Names Over Time

The tables below shows the ten most popular given male names about London c. 1120 to 1994, and the ten most popular female names in North England (Yorkshire, Cumbria, Northumberland) c. 1350 to 1994. The years given are approximate birth years. The average year of birth was estimated relative to the date of the compilation and the probable ages of the persons in the compilation. The data come from a variety of sources, which have used different (and often not explicit described) approaches to standardizing and grouping names.

Scholars interested in additional name lists should consult Smith-Bannister (1997), Appendix C. That appendix lists, at decade intervals, the fifty most popular male and female names in forty English parishes from 1538-49 to 1690-1700. Unfortunately, the frequency of specific names are not given, nor are sample sizes. The weights given to individual parishes in each decade sample apparently change, but details are not given.

**Ten Most Popular Male Names in London**

| Rank | Name | Year c. 1120 | Name | Year c. 1260 | Name | Year c. 1510 |
|---|---|---|---|---|---|---|
| 1 | Willelm | 6.6% | John | 17.6% | John | 24.4% |
| 2 | Robert | 5.0% | William | 14.4% | Thomas | 13.3% |
| 3 | Ricard | 4.2% | Robert | 7.7% | William | 11.7% |
| 4 | Radulf | 3.6% | Richard | 7.0% | Richard | 7.3% |
| 5 | Roger | 3.2% | Thomas | 5.3% | Robert | 5.6% |
| 6 | Herbert | 2.2% | Walter | 4.4% | Ralph | 3.3% |
| 7 | Hugo | 1.8% | Henry | 4.1% | Edward | 3.0% |
| 8 | Johannes | 1.3% | Adam | 3.1% | George | 2.1% |
| 9 | Anschetill | 1.1% | Roger | 2.9% | James | 1.9% |
| 10 | Drogo | 1.1% | Stephen | 2.3% | Edmund | 1.6% |
|  | Sample Size | 912 | Sample Size | 814 | Sample Size | 427 |

| Rank | Name | Year c. 1610 | Name | Year c. 1825 | Name | Year c. 1994 |
|---|---|---|---|---|---|---|
| 1 | John | 21.0% | William | 16.3% | Jack | 3.2% |
| 2 | William | 11.4% | John | 13.5% | James | 3.1% |
| 3 | Thomas | 11.4% | George | 9.4% | Daniel | 2.2% |
| 4 | Richard | 5.2% | James | 8.6% | Thomas | 2.2% |
| 5 | Samuel | 5.0% | Thomas | 8.6% | Michael | 1.6% |
| 6 | Henry | 4.8% | Henry | 7.6% | Alexander | 1.5% |
| 7 | Edward | 4.5% | Charles | 5.8% | Matthew | 1.4% |
| 8 | James | 3.5% | Joseph | 3.7% | Luke | 1.3% |
| 9 | Joseph | 2.6% | Edward | 3.5% | Samuel | 1.3% |
| 10 | Robert | 2.4% | Robert | 3.1% | George | 1.3% |
|  | Sample Size | 463 | Sample Size | 48275 | Sample Size | 51097 |

Sources and Notes:

The list for 1120 was compiled from the Winton Doomesday book for the year 1148. The list is given in Barlow et al. (1976), Table 8, p. 187. No details are provided regarding any name standardization done.

The list for 1260 was compiled from the London Subsidy Rolls of 1292. The list is given in Ekwall (1951) p. 35. Variants included under the given name heading are as follows: John (Jon), Walter (Water), and Henry (Hanry, Hary, Herri).

The list for 1510 was compiled from baptismal names in five London parishes, 1540-1549. The parishes are: St. Peter's upon Cornhill; St. De'nis Backchurch; Christ Church Newgate; Kensington; St. Antholin, Budge Row. All baptismal names, as recorded in Harleian Society Publications, were included in the compilation. The list is given in Stewart (1948), Table 1, p. 110. The source notes: "I have tried to ignore mere variations of spelling, but to count separately the different forms of the same name, such as Henry and Harry, Augustine and Austin. … Spellings have been normalized to conform with those of the King James Bible, or with modern usage for non-Biblical names." See Stewart (1948), footnote pp. 109-10.

The list for 1610 was compiled on the same basis as the 1510 list, but for the years 1640-1649. See Stewart (1948) Table 2, p. 112.

The 1825 list is from persons born between 1819 and 1830 in London, and still alive and recorded in the U.K. Census of 1881. The complete Census of 1881 is available from the Genealogical Society of Utah (1997). Names given in the Census were standardized using GINAP v. 1.0 name standardization, available at http://users.erols.com/dgalbi/names/ginap.htm.

The list for 1994 includes all males born in Greater London in 1994 who registered with the National Health Service Central Register. See Merry (1995), Table 21. The source lists the top 50 names. From that compilation, the list given above groups Jamie with James and Jake with Jack.

**Ten Most Popular Female Names in North England**

| Rank | Name | Year c. 1350 | Name | Year c. 1530 | Name | Year c. 1640 |
|---|---|---|---|---|---|---|
| 1 | Alicia | 22.4% | Jane | 16.0% | Elizabet | 16.6% |
| 2 | Agnete | 14.9% | Elizabet | 14.7% | Ann | 16.4% |
| 3 | Johanna | 13.2% | Margaret | 14.2% | Jane | 13.1% |
| 4 | Emma | 7.7% | Agnes | 8.2% | Margaret | 12.8% |
| 5 | Elena | 5.4% | Isabella | 7.9% | Mary | 9.9% |
| 6 | Isabella | 5.3% | Alice | 7.7% | Isabella | 5.9% |
| 7 | Margareta | 5.1% | Ann | 7.3% | Ellen | 3.6% |
| 8 | Cecilia | 4.5% | Ellen | 3.6% | Alice | 3.6% |
| 9 | Matilda | 3.6% | Catherin | 2.8% | Dorothy | 2.9% |
| 10 | Juliana | 3.3% | Mary | 1.9% | Frances | 1.9% |
|  | Sample Size | 1794 | Sample Size | 852 | Sample Size | 2888 |

| Rank | Name | Year c. 1730 | Name | Year c. 1825 | Name | Year c. 1994 |
|---|---|---|---|---|---|---|
| 1 | Ann | 18.1% | Mary | 20.3% | Rebecca | 3.7% |
| 2 | Mary | 16.7% | Jane | 13.5% | Lauren | 3.6% |
| 3 | Elizabeth | 15.3% | Elizabeth | 12.9% | Amy | 2.8% |
| 4 | Jane | 11.2% | Ann | 11.6% | Laura | 2.6% |
| 5 | Margaret | 10.6% | Margaret | 9.4% | Jessica | 2.6% |
| 6 | Isabella | 4.8% | Isabella | 6.2% | Rachel | 2.4% |
| 7 | Hannah | 3.0% | Sarah | 5.8% | Sophie | 2.4% |
| 8 | Ellen | 2.9% | Hannah | 3.6% | Hannah | 2.4% |
| 9 | Dorothy | 2.6% | Ellen | 3.3% | Sarah | 2.3% |
| 10 | Sarah | 2.1% | Catherin | 1.6% | Emma | 2.0% |
|  | Sample Size | 2038 | Sample Size | 24857 | Sample Size | 17719 |

Sources and Notes:

The list for 1350 is from the assessment roll of the 1379 Poll Tax for Howdenshire Hundred in Yorkshire East Riding. See Gwynek (n.d.). The names were listed in Latinized forms in the ablative case. Variants and dimunitives were combined as follows: Alicia (Alisia, 1), Agnete (Augnete, 1; Annya, 1), Emma (Emmota, 7), Margareta (Mergareeta, 1; Marg, 1; Marg…, 1; Mar', 1; Magota, 6), Cecilia (Cicilia, 19; Sissota, 3; Syssota, 1).

The sources for 1530, 1640, and 1730 are from parish marriage registers in Northumberland and Durham for the years 1538-68, 1650-70, and 1740-60, respectively. For complete standardized given name list, see Galbi (2001). Names were standardized using GINAP v. 1.0 name standardization, available at
http://users.erols.com/dgalbi/names/ginap.htm.

The 1825 list is from persons born between 1819 and 1830 in Cumberland, Durham, and Northumberland, and still alive and recorded in the U.K. Census of 1881. The complete Census of 1881 is available from the Genealogical Society of Utah (1997). Names given in the Census were standardized using GINAP v. 1.0

The list for 1994 includes all females born in Cumbria, Durham, Cleveland, Tyne & Wear, and Northumberland in 1994 who registered with the National Health Service Central Register. See Merry (1995), Table 5. The source lists the top 50 names. From that compilation, the list given above grouped Rachel and Rachael.

## Additional References

[1] The title of this paper originally was "Long-Term Trends in Personal Given Name Frequencies in the UK." The title has been changed to more accurately indicate the geography of the name samples considered.

[2] The opinions and conclusions expressed in this paper are those of the author. They do not necessarily reflect the views of the Federal Communications Commission, its Commissioners, or any staff other than the author. Author's address: dgalbi@fcc.gov; FCC, 445 12'th St. SW, Washington, DC 20554, USA.

[3] See the GINAP webpage, http://users.erols.com/dgalbi/names/ginap.htm. The principle for coding is to group together names that either sound the same, have the same public meaning, or changed only in the recording process (spelling errors, recording errors, etc.).

[4] Note that name standardization helps to control for changes in names used as a person grows older, e.g. a correlation between nicknames or informal names and age. Thus name standardization is particularly important in analyzing time trends when the data come from naming cohorts constructed by age. That is the case for this paper's data on nineteenth century names.

[5] Clark (1992) pp. 552, 558-562. There is no evidence that Norman clergy or royal officials compelled the English to adopt Norman names.

[6] This is true because $\log(a/b)=\log(a)+\log(b)$. The logarithm of name frequency differs from the logarithm of name popularity only by an additive factor. Name popularity rank and name frequency rank are of course identical.

[7] Gabaix (1999) shows that, when the appearance rate for new cities is not too high, it has no effect on the slope of the power law describing city sizes. If the appearance rate for new cities rises above a certain threshold, than the slope depends on the appearance rate. Cities can be analogized to name types.

[8] For population and income statistics for 1700 and earlier, see Mayhew (1995) Table I, and Snooks (1995), Table 3.5. For current population statistics, see UK National Statistics, Key Population and Vital Statistics, online at http://www.statistics.gov.uk/statbase/Product.asp?vlnk=539&More=N . The large changes in the structure of the economy over the past two hundred years make estimating changes in per capita income subject to significant uncertainty. The figure of 100 is my estimate based on my knowledge of the economic history literature.

[9] See AGNAMES webpage at http://users.erols.com/dgalbi/names/agnames.htm